# Effect of Injector Geometry in Breakup of Liquid Jet in Crossflow – Insights from POD

Anubhav Sinha*
*Indian Institute of Technology (Banaras Hindu University), Varanasi -221005, India*

The present study aims to investigate the role of injector geometry, particularly injector tube length to diameter ratio (*L/D*) in liquid jet stability and breakup in presence of crossflow. Water is injected into a crossflow of air. Aerodynamic Weber number (*We*) and liquid Reynolds number ($Re_l$) are systematically varied to observe various breakup modes. Bag breakup and surface stripping is observed for different operating conditions. Time-resolved jet trajectory images are processed using Proper Orthogonal Decomposition (POD) algorithm. POD mode shapes and corresponding Power Spectral Density (PSD) plots are analyzed to study breakup process and probe role of injector geometry effects. Further, high-resolution images are captured for the near-nozzle region. Detailed comparison is made for various cases. It is observed that with increase in (*L/D*), the jet surface becomes more turbulent and unstable. This results in early breakup and lower jet penetration

## Nomenclature

| | | |
|---|---|---|
| *L* | = | injector tube length |
| *D* | = | injector/ jet diameter |
| *Re* | = | Reynolds number |
| *We* | = | Aerodynamic Weber number |
| *q* | = | Momentum ratio of jet and crossflow |
| *U* | = | velocity |
| *ρ* | = | density |

*Assistant Professor, Department of Mechanical Engineering, IIT (BHU), er.anubhav@gmail.com

| $\sigma$ | = | surface tension |
| $u'_{rms}$ | = | RMS turbulence fluctuations in velocity |

*Subscripts*

| $g$ | = | gas phase |
| $l$ | = | liquid phase |

## I. Introduction

LIQUID Jet in crossflow (LJIC) is an important configuration for fuel injection in aerospace applications, particularly in gas turbine engines, afterburners, ramjets, etc. Liquid jet is injected into a crossflow of high velocity airflow. The jet disintegrates into ligaments and drops. The breakup is understood to be caused by aerodynamic forces. However, liquid turbulence and jet instability also aid the breakup process. Ligaments and drops formed from primary breakup further undergo secondary breakup to produce smaller droplets which eventually evaporate, mix, and combust. Jet breakup and droplet formation impacts fuel mixing, combustion efficiency, pollutant formation, etc. It could also play a major role in thermo-acoustic instabilities, and their control. Hence, it becomes important to understand the breakup process and factors which govern it. As the jet enters the crossflow, aerodynamic forces start acting on the jet surface. Aerodynamic Weber number (*We*) is an important parameter for LJIC configuration, which is defined as:

$$We = \frac{\rho_g U_g^2 D}{\sigma} \tag{1}$$

For lower *We*, the column bends, gets flattened, and breaks undergoing column or bag breakup. For larger *We*, surface stripping is observed, and jet undergoes a more rapid breakup. Wu et al. [1, 2] carried out a detailed investigation of LJIC for a wide range of operating conditions and classified different regimes of jet breakup. Their regime map is based on *We* and jet to crossflow momentum ratio (*q*), where *q* is defined as:

$$q = \frac{\rho_l U_l^2}{\rho_g U_g^2} \tag{2}$$

Subscript *l* refers to the liquid jet and *g* refers to the crossflow air. The injector used Wu et al. [1, 2] had a fixed (*L/D*) ratio of 4. Mazallon et al. [3] investigated LJIC using different jet liquids and operating conditions. They conclude *We* to be the controlling parameter in jet breakup, and relatively minor effect from *q*. Another study was conducted by Sallam et al. [4]. They have also studied different breakup regimes for nonturbulent jets and plotted their observation in a regime map. They concur that *We* is the determining factor for laminar jets and *q* or $Re_l$ has no major impact on jet breakup. Ng et al. [5] conducted a detailed experimental study characterizing the column and bag breakup regimes. They used a super-cavitating nozzle to generate a nonturbulent jet. They have analyzed the bag formation and its rupture. The jet instability is attributed to be Rayleigh Taylor instability. They have derived relations for surface waves and found a good match with their experimental observation. Lee et al. [6] have undertaken an experimental study of turbulent jet breakup in presence of crossflow. They focused on the role of liquid turbulence in jet breakup, in contrast to previous studies which focused on aerodynamic effects. A general agreement is that increasing air velocity (resulting in higher *We*) will enhance the breakup. However, Lee et al. [6] suggested that the breakup is primarily caused by jet turbulence, and it will get suppressed by increasing air velocity, in the wind-ward side. In another study, Sallam et al. [7] have carried out experiments on both turbulent and nonturbulent jets and compared breakup regimes. They have investigated the role of jet turbulence in jet disintegration. They have introduced a new non-dimensional number, Faeth number to characterize breakup of turbulent jets. They have argued that in the turbulent breakup regime, turbulent eddies in jet are primary responsible for breakup. If the crossflow air velocity is increased, the pressure force on jet will increase and will suppress breakup.

It is interesting to note that the injector used by Mazallon et al. [3] and Sallam et al. [4, 7] focused on nonturbulent jet and have used an injector with lower *L/D* (*L/D*<3), while Lee et al. [6] and Sallam et al. [7] have used an injector with (*L/D*)>100 and generated a turbulent jet. It is important to note that this understanding was there that (*L/D*) impacts jet stability and transition to turbulence [8, 9], but no comprehensive study was undertaken to explore this aspect in LJIC. Madhabushi et al. [10] have carried out a review of previous studies on jet breakup regimes. They have highlighted that the momentum ratio (*q*) can be replaced by liquid Reynolds number ($Re_l$) in regime maps and emphasized that $Re_l$ offers more physical insight, as jet transition from laminar to turbulent can be captured by the value of $Re_l$. It is also highlighted that the effect of injector geometry has been overlooked. Brown and McDonell [11] have captured near-nozzle images of jet breakup in crossflow. They observed jet trajectories captured in their experiments in good agreement with correlations available in the literature [2, 12] . Interestingly, they have admitted that they get a good match because they have similar injector geometry that was used to derive those correlations. They have also suggested that the role of injector geometry needs to be studied in detail.

Due to practical significance of LJIC, this configuration has been a subject of active research since decades [13, 14]. However, there is no general agreement on something as fundamental as regime map for breakup or the parameters that govern breakup [13]. The boundaries between different regimes varies for each study, depending on their test conditions and injector geometry used [10]. Even for jet trajectory and penetration, there are numerous correlations available in the literature focusing on different parameters [14, 15]. This could partly be attributed to the neglecting the effect of injector geometry in previous literature. Injector geometry is considered so insignificant, several papers haven't even provided all the geometry details regarding their injector [10]. A recent study [16] investigated the effect of L/D on LJIC. However, that study entirely focused on wave growth on jet surface, and jet breakup was not investigated. There are some experimental investigations [17, 18] which focus on the role of injector geometry on generation of surface waves on liquid jets. They have characterized wavelength and studied transition to turbulence for a liquid jet discharged into a quiescent ambient. Portillo et al. [18] have attributed the transition to turbulence to the instability near the nozzle exit. However, this aspect remains largely unexplored for LJIC configuration.

LJIC configuration has also been investigated using computational tools, where several important physical processes are captured in numerical simulations. Behzad et el. [19] have carried out a numerical study of LJIC for high pressure conditions. They have captured surface waves and droplet stripping from the jet at high $We$ cases. They have kept $Re_l$ low to obtain a laminar jet. Herrmann [20] has carried out numerical simulations for the conditions used by [11] and is able to capture jet primary breakup characteristics for a turbulent jet. In this study, injector internal flow is also simulated, and injector induced instability is captured, which is generally neglected in other numerical studies [21, 22].

Another important point to remember is that breakup in JICF configuration is a highly unsteady process. It has been pointed by various researchers [10, 11] that a single instantaneous image per case is not sufficient to investigate various breakup modes, their associated structures, and frequencies. Time-resolved videos or a large set of images are generally needed to understand the dynamics of this configuration. Time resolution for image capture is determined by the highest frequency observed in the flow field, which typically leads to thousands of frames per second. Hence, in order to understand the breakup process for one case, it is required to analyze a large set of instantaneous images. Another alternative is to process these images using Proper Orthogonal Decomposition (POD) analysis. POD is a data reduction technique used to obtain low-dimensional approximations of high-dimensional data. Berkooz et al. [23] have explained the basics of POD analysis with its applications. Liang et al. [24] and Taira et al. [25, 26] have also detailed the basic mathematical principles and applications of POD. The present work uses snapshot POD, which is due to Sirovich [27]. Taira et al. [25, 26] have compiled a useful reference for POD and other modal reduction techniques, along with applications from fluid mechanics. POD is particularly useful for identifying coherent structures in fluid flow and their associated frequencies. POD was recently used to investigate airblast spray dispersion in crossflow [28], and a regime map was proposed based on POD analysis. However, POD has not been utilized to uncover the dynamics of JICF configuration. Arienti, and Soteriou [29] claim to have the only paper where POD is applied to study JICF configuration.

To summarize, there is a need to identify the role of injector geometry in jet breakup in crossflow. It requires both spatial and time resolved imaging to gain complete understanding of the complex physics of jet breakup. Also, POD is a tool which can help in identifying hidden structures in fluid flow problems, and not been fully utilized in JICF configuration. This paper attempts to address this issue.

## II. Experimental Details

### A. Experimental set-up

The experimental set-up consists of an air supply system for the crossflow air, and a liquid supply system for the injector. Schematic of the facility shown in Fig. 1. The air supply is a blowdown type of wind tunnel. Air is stored in tanks at a high pressure. After being released form the tanks, air passes through a series of pipelines, control valves, etc. to the diverging section to a settling chamber, and finally through a converging section to the test section. The converging section is designed following the guidelines of Mehta and Bradshaw [30] and Brassard and Ferchichi [31]. The velocity profile and turbulence levels are characterized using a hot wire anemometer at the entry of the test section. The velocity profile is found to be fairly uniform with the velocity values varying within 2.5% of the mean. The turbulence intensity ($u'_{rms}/U$) is measured to be around 3%. Details about the velocity profile and turbulence measurement are provided elsewhere [32]. The test section is a rectangular duct of 50X54 mm cross section and 250 mm length. It is made optically accessible by quartz side walls. The injector is flush fitted to the center of the bottom wall. The injector diameter (*D*) is fixed to 0.5 mm and the length (*L*) is varied to obtain (*L/D*) ratios of 10 and 100. Liquid flow rate to the injector is controlled by a Coriolis-based mass flow controller (Make: Bronkhorst, Model: mini-Cori-Flow, Accuracy: ±0.2% of flow rate).

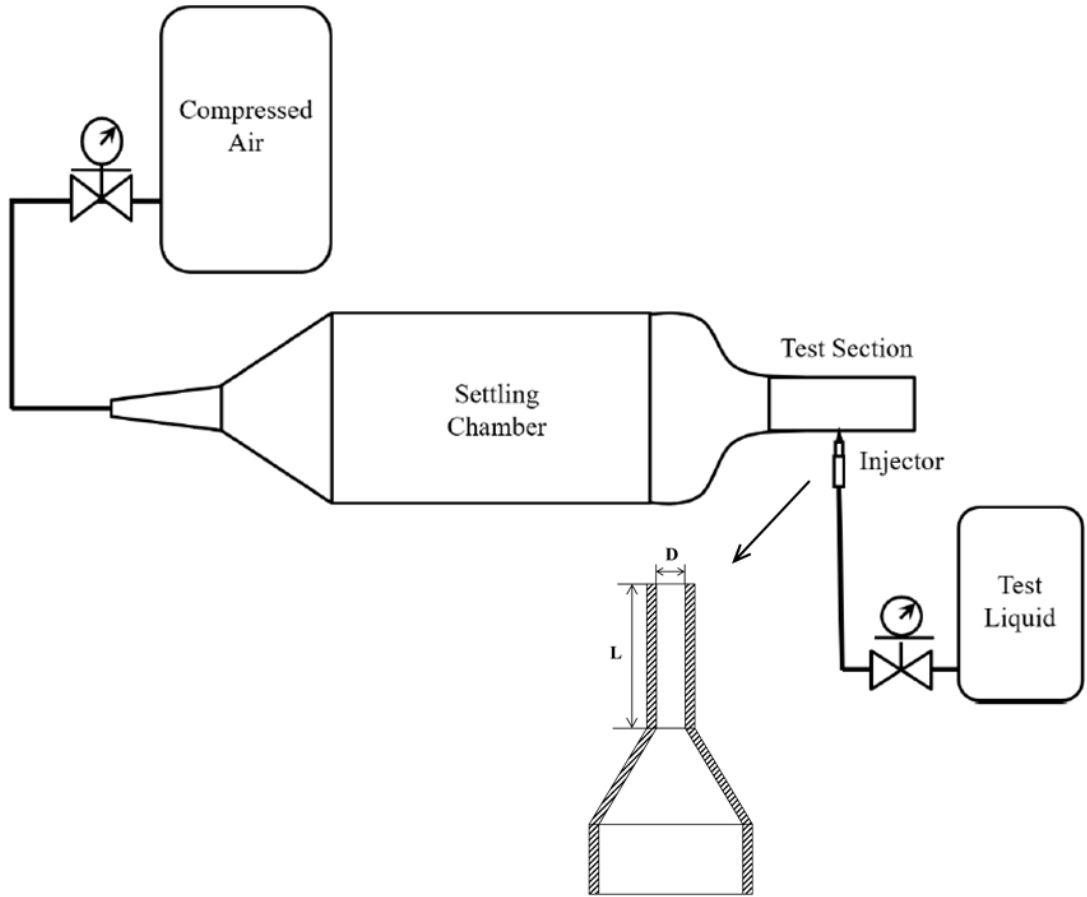

**Fig. 1. Schematic of Experimental facility. Injector geometry details are shown in an enlarged image.**

**B. Imaging systems**

Two types of imaging systems are used in the present study: near nozzle spatially resolved imaging, and time-resolved jet trajectory imaging. Near-nozzle images are captured using a 4MP CCD camera (model: LaVison ImagerProX4M). Pulsed laser (model: Nano-L-200-15-PIV) is used in this system. The laser pulse is around 10 nano seconds. However, laser falls on a diffuser plate with Rhodamine dye, which fluoresces for around 20 nano seconds. This diffuser plate is used for background illumination and gives the effective exposure time of 20 nano seconds for the CCD camera. Due to superior background illumination from the laser system and such small exposure time, the images obtained appear to be frozen in time, and show detail jet breakup morphology. This system captures images at 120 pixels per mm for the settings used in this study. Although, this imaging system provides a very high spatial resolution, the temporal resolution is limited by the Laser repetition rate, which is 15 Hz. For temporally resolved images of jet trajectory and breakup, a high-speed CMOS camera (model: Photron Fastcam SA5) is used in conjugation with a DC powered 500W halogen lamp. Images are captured at 10,000 frames per second, with an exposure time of 1 microsecond, and spatial resolution as 30 pixels/ mm.

| Case | Liquid Reynolds number ($Re_l$) | Aerodynamic Weber Number ($We$) | Momentum Ratio ($q$) | (L/D) |
|---|---|---|---|---|
| J1 | 2545 | 16 | 11 | 10 |
| J2 | 2545 | 16 | 11 | 100 |
| J3 | 5090 | 16 | 45 | 10 |
| J4 | 5090 | 16 | 45 | 100 |
| J5 | 10180 | 79 | 36 | 10 |
| J6 | 10180 | 79 | 36 | 100 |

**Table 1. Test conditions of the cases considered in this study.**

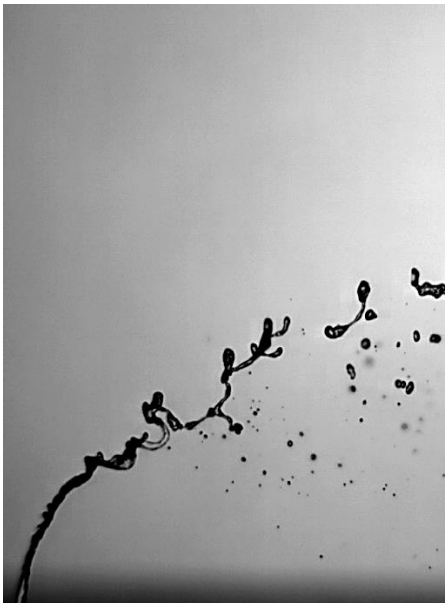
(i) Case J1

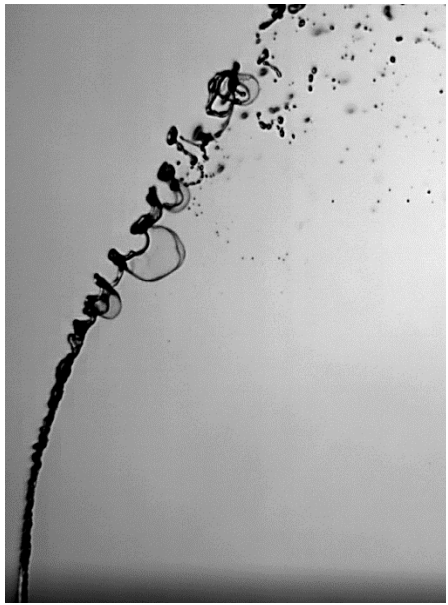
(iii) Case J3

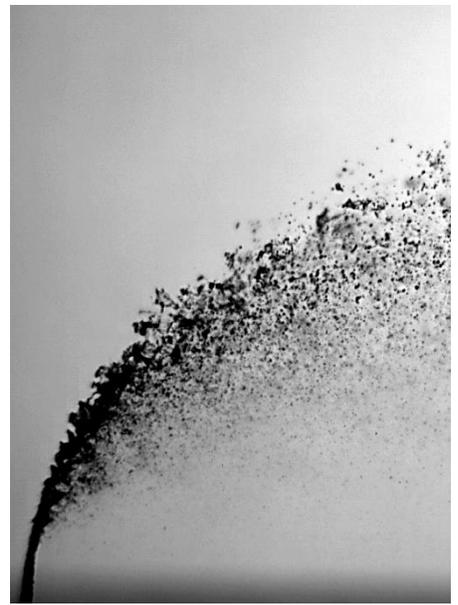
(v) Case J5

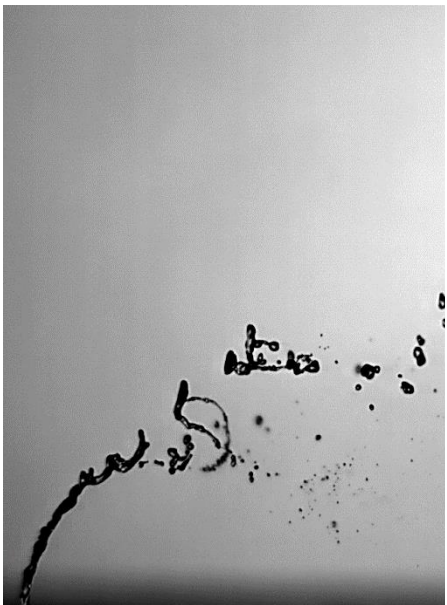
(ii) Case J2

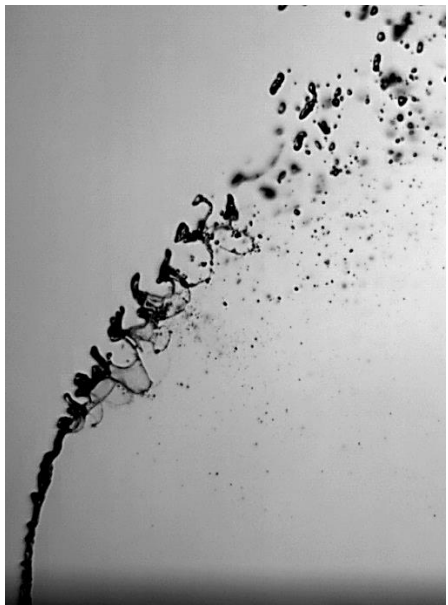
(iv) Case J4

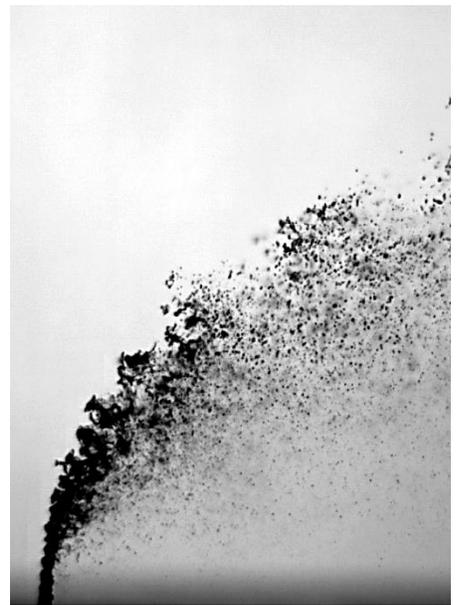
(Vi) Case J6

**Fig. 2    Instantaneous images captured using CMOS camera for various cases considered in this study.**

**C. Experimental Conditions**

The test conditions for various cases investigated in this study is summarized in Table 1. For each pair of cases, the operating conditions are maintained same, while (*L/D*) changes from 10 to 100. Liquid Reynolds number ($Re_l$) is kept under laminar conditions for the first pair (J1 and J2), near critical value for the second pair (J3 and J4) and in turbulent regime for third pair (J5 and J6). The critical value of $Re_l$ is defined as 5000 by Madhabushi et al. [10] or laminar to turbulence transition. First two pairs are at low *We* of 16 and last pair is at a relatively higher *We* of 79. Momentum ratio is varied from 11 to 45. Typical instantaneous high-speed images for each case is shown in Fig. 2. These images are used for POD processing discussed in later sections.

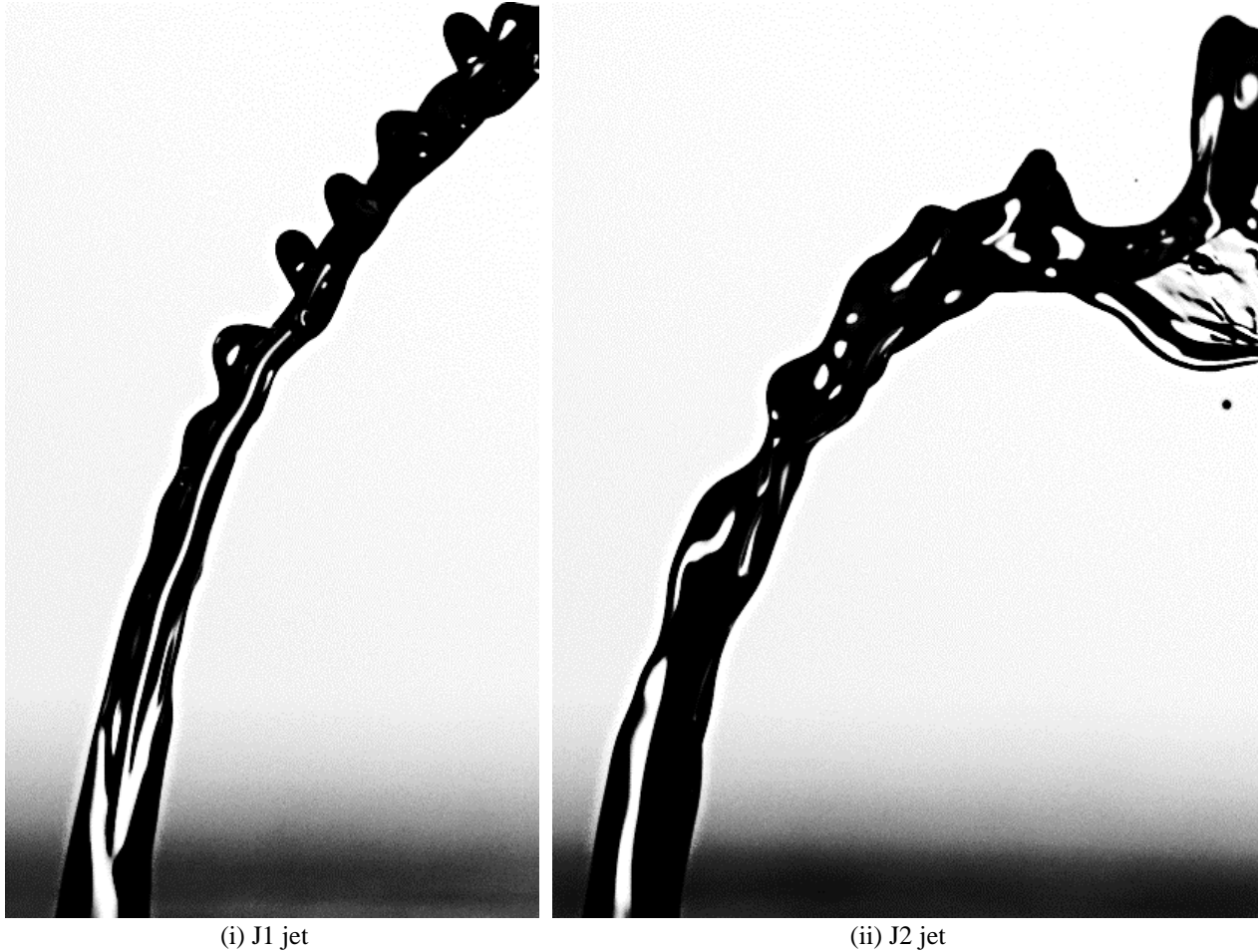

       (i) J1 jet        (ii) J2 jet

**Fig. 3   Near-nozzle jet structure for J1 and J2 cases.**

## III. Results and Discussion

Figure 3. shows the near-nozzle images for J1 and J2. Both these cases have the same operating conditions, except for the injector geometry (*L/D*). Both jets look laminar (or glassy) as they emerge from the injector. However, J1 appears to be more stable, while J2 deflects more and breaks more rapidly. J1 exhibits surface waves of smaller wavelength and amplitude while J2 shows waves of larger wavelength and amplitude. It is interesting to note that the waves in J1 are visible in wind-ward side only, and not so apparent in the lee-ward side. Whereas in J2, the waves are prominent in both wind-ward as well as lee-ward sides of the jet and aids in early breakup. Jet flattening for bag formation is also more prominent in J2.

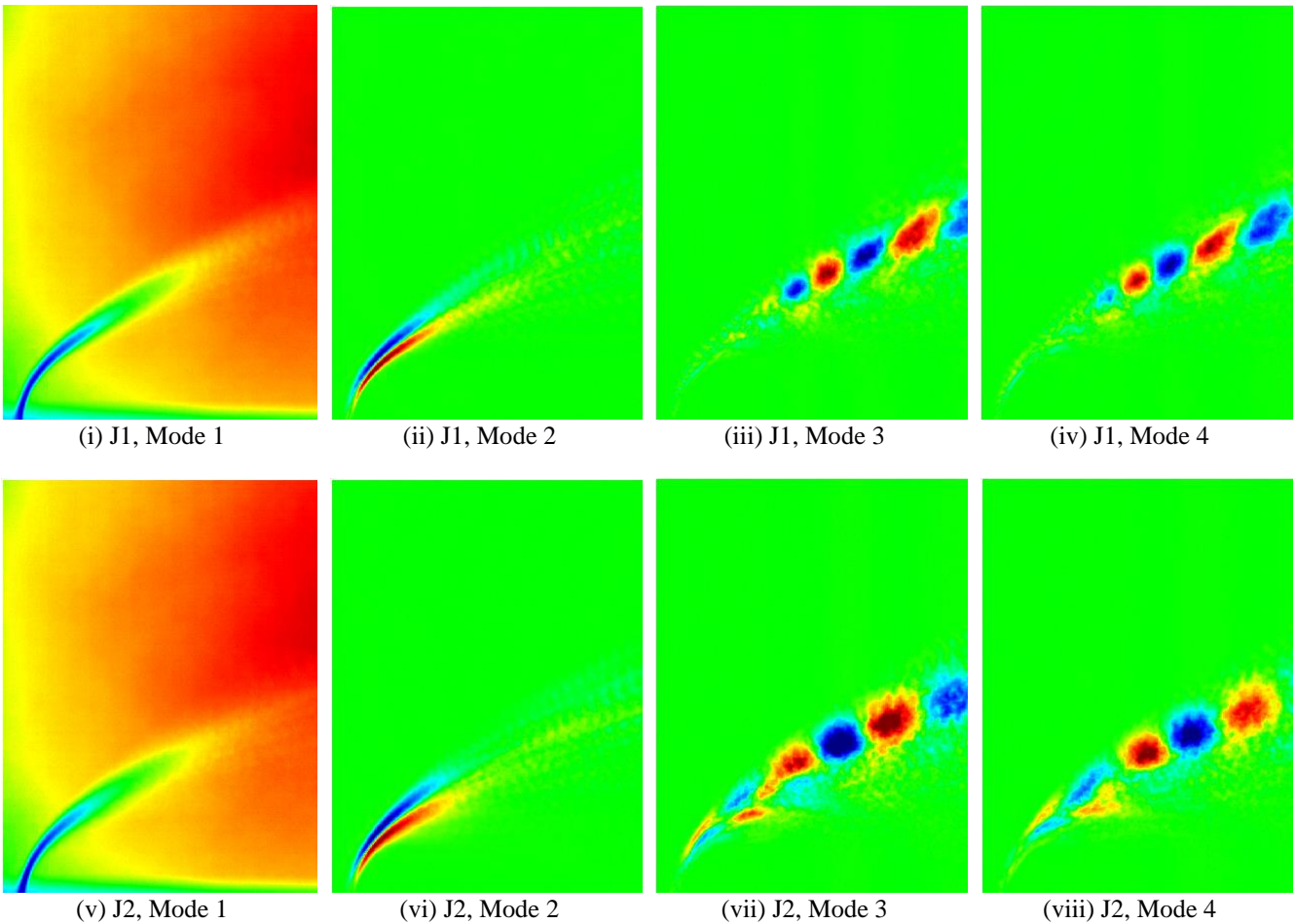

(i) J1, Mode 1    (ii) J1, Mode 2    (iii) J1, Mode 3    (iv) J1, Mode 4

(v) J2, Mode 1    (vi) J2, Mode 2    (vii) J2, Mode 3    (viii) J2, Mode 4

**Fig. 4    POD Mode shapes for J1 and J2 cases.**

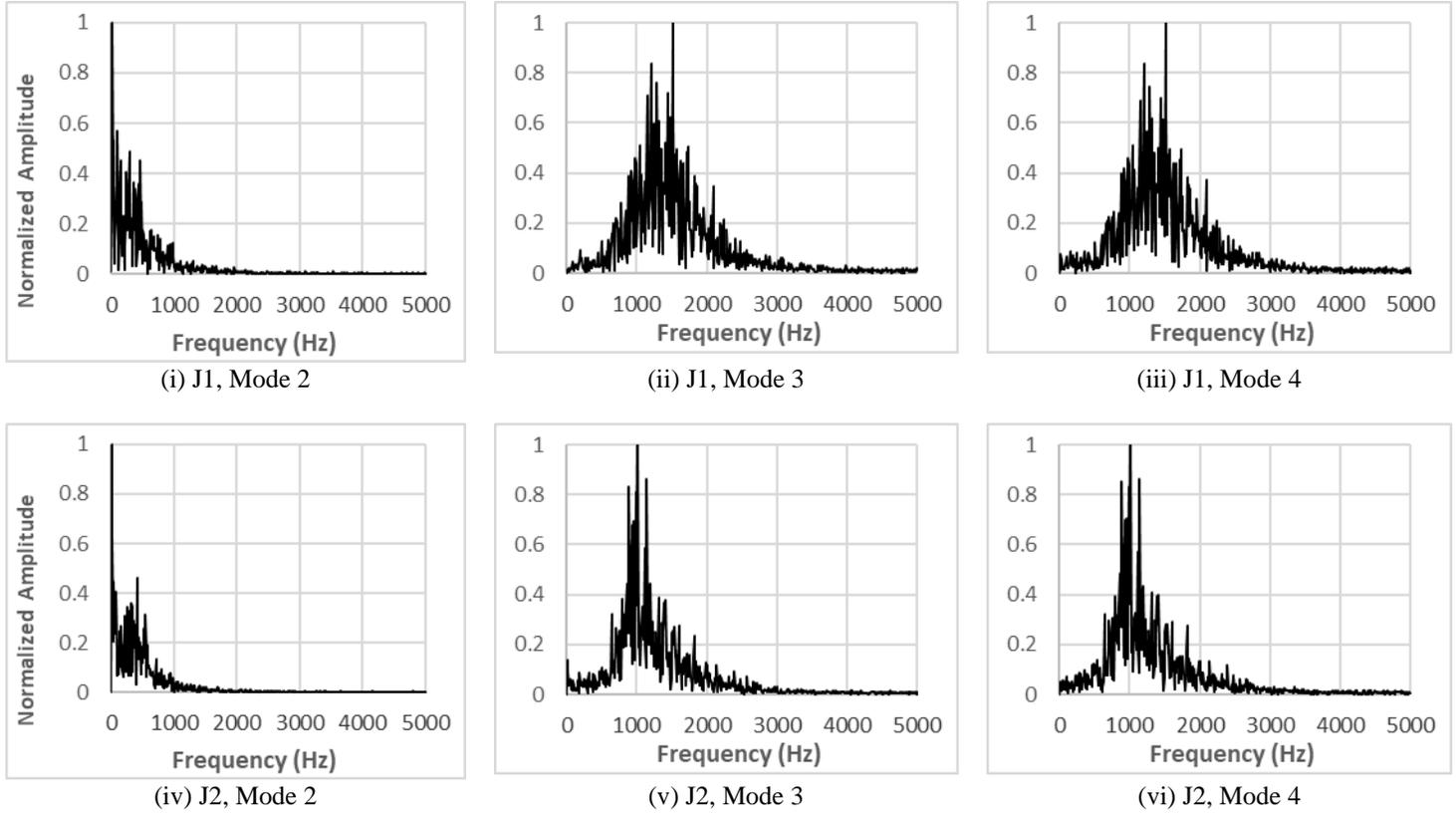

Fig. 5    PSD spectrum for J1 and J2 cases

To further investigate the underlying process, POD analysis is carried over a large number of time-resolved images. Figure 4 shows the POD modes for J1 and J2 and Fig. 5 shows the corresponding PSD modes, highlighting dominant frequencies. The first POD mode is the average mode, which shows the mean jet trajectory. Since it is the average mode, its corresponding PSD is not shown. The mean trajectory of J1 and J2 looks similar, with J1 having a slightly higher penetration. However, the difference gets highlighted in higher modes. Mode 2 captures the jet oscillation about its mean position. It takes place at a peak frequency of around 500 Hz for both the cases. However, jet oscillation in J2 is more pronounced as compared to J1. Jet oscillation is an important motion in LJIC and is observed for all the cases considered in this study. Generally, jet oscillates at around 500 Hz for all the cases implying it to be a fundamental jet motion. POD mode 3 and 4 are similar and denote bag/ ligament breakup and transport along the jet trajectory. Mode 3 and 4 for J1 show a peak around 1500 Hz, whereas a peak of around 1000 Hz is observed for case J2. Bags and ligaments formed in J2 are larger than those formed in J1. Third and fourth modes of J2 also show signatures of column oscillation. This stresses on the fact that the J2 jet is more unstable and jet oscillation is happening at multiple frequencies, highlighting the complexity of the breakup process.

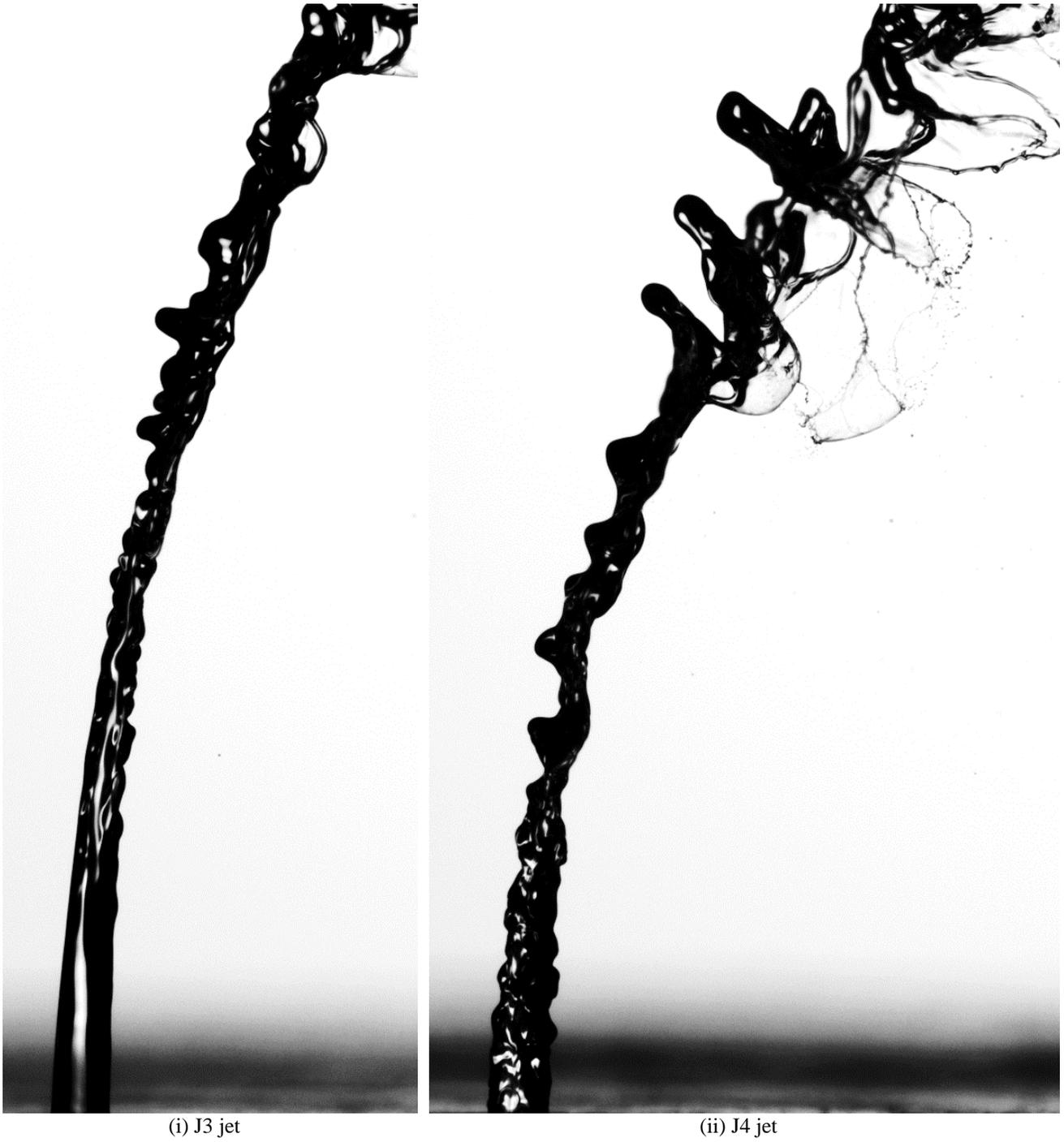

(i) J3 jet  (ii) J4 jet

**Fig. 6. Near-nozzle jet structure for J3 and J4 cases.**

Case J3 and J4 maintain the same Weber number as J1 and J2 cases. However, the jet Reynolds number and momentum ratio has increased. Figure 6 presents the near nozzle images for these cases. J3 jet emerges from the injector with a glassy appearance indicating laminar jet, and later gets destabilized by disturbances both on lee-ward and wind-ward sides. However, J4 jet is turbulent as soon as it emerges from the injector. This leads to a highly irregular surface, large-scale distortions, and rapid breakup. It must be emphasized that this drastic change in breakup behavior can entirely be attributed to nozzle geometry, as all other operating parameters are maintained same in J3 and J4. These cases are further probed using POD analysis and POD mode shapes and PSD plots are shown in Fig. 7 and 8 respectively.

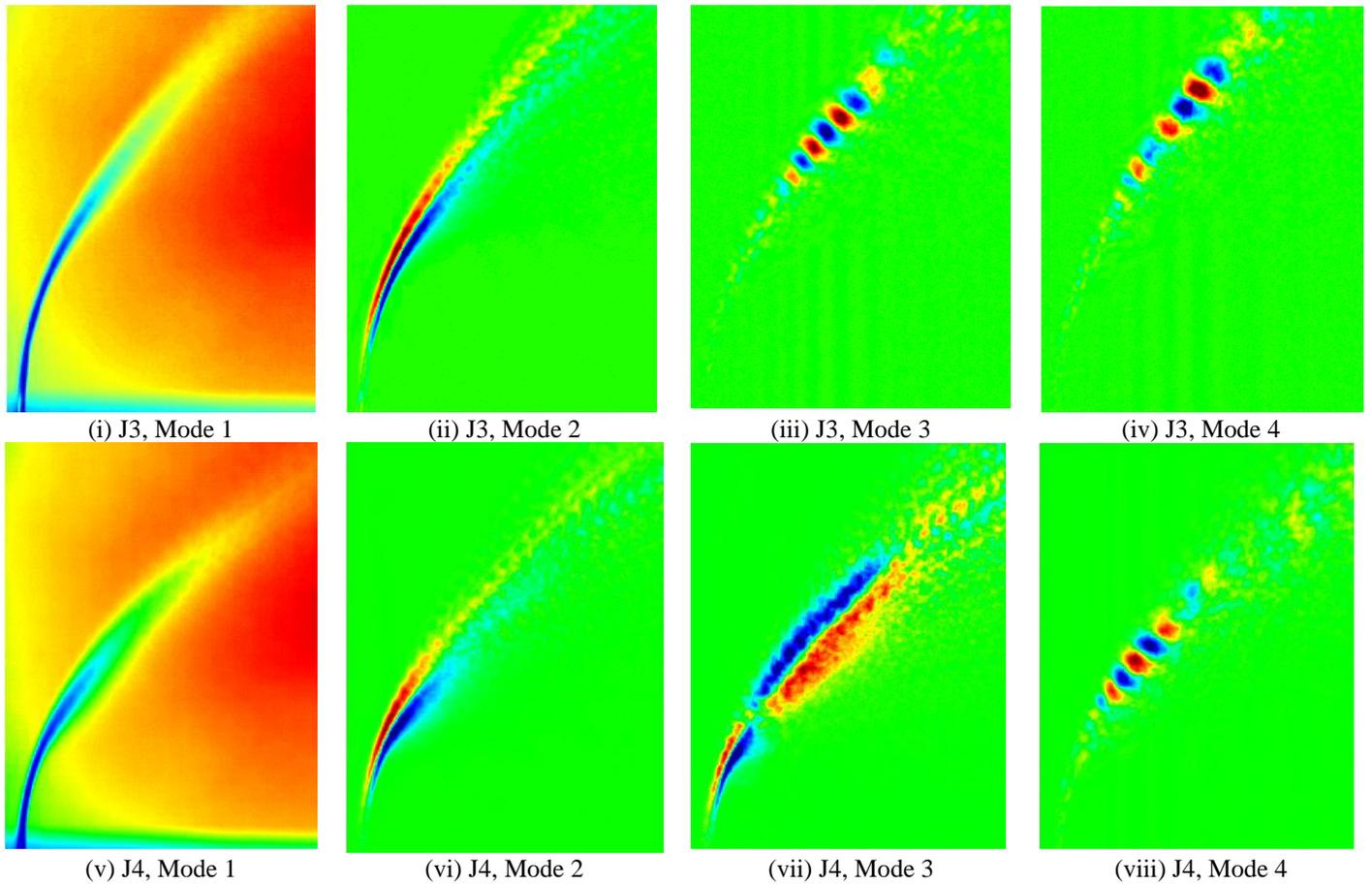

**Fig. 7. POD Mode shapes for J3 and J4 cases**

From Fig. 7, it is observed that the mean jet trajectory of J4 is slightly shorter than J3 and is more deflected in the crossflow direction. Column oscillation is more prominent in J4 as compared to J3 and has a wider PSD spectrum as compared to J3. Column oscillation for J3 has a prominent peak around 500 Hz. Mode 3 and 4 for J3 denote the ligament/ bag breakup and their transport. This ligament breakup happens at a broadband spectrum of around 4000 Hz, as observed in Fig. 8. On the other hand, mode 3 of J4 shows jet flapping which is a complex form of jet oscillation and occurs around the same frequency. It signifies more complex and violent breakup behavior of J4. Bag breakup is captured in the fourth POD mode of J4, and this also shows a broadband spectrum between 3500 to 4500 Hz. PSD showing a broad range of frequencies indicate more complex process involving multiple timescales.

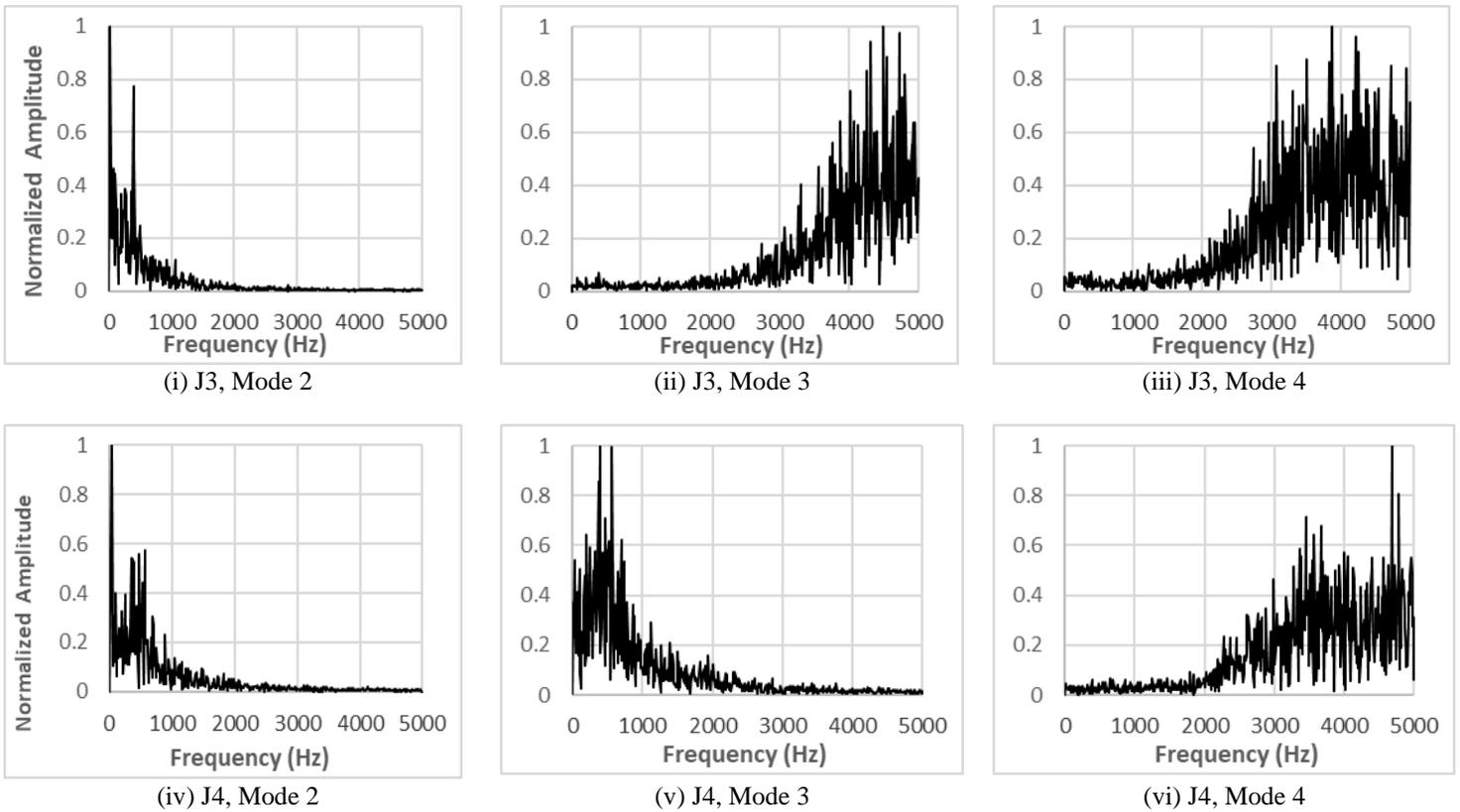

Fig. 8    PSD spectrum for J3 and J4 cases

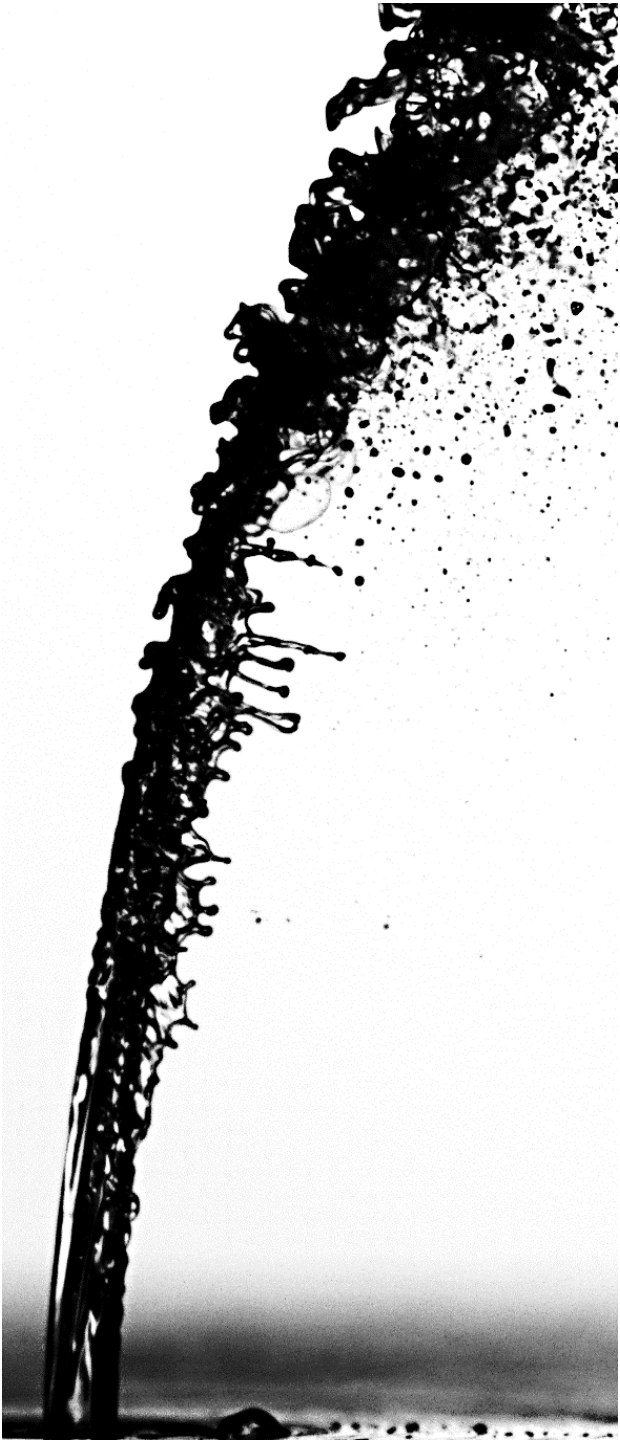 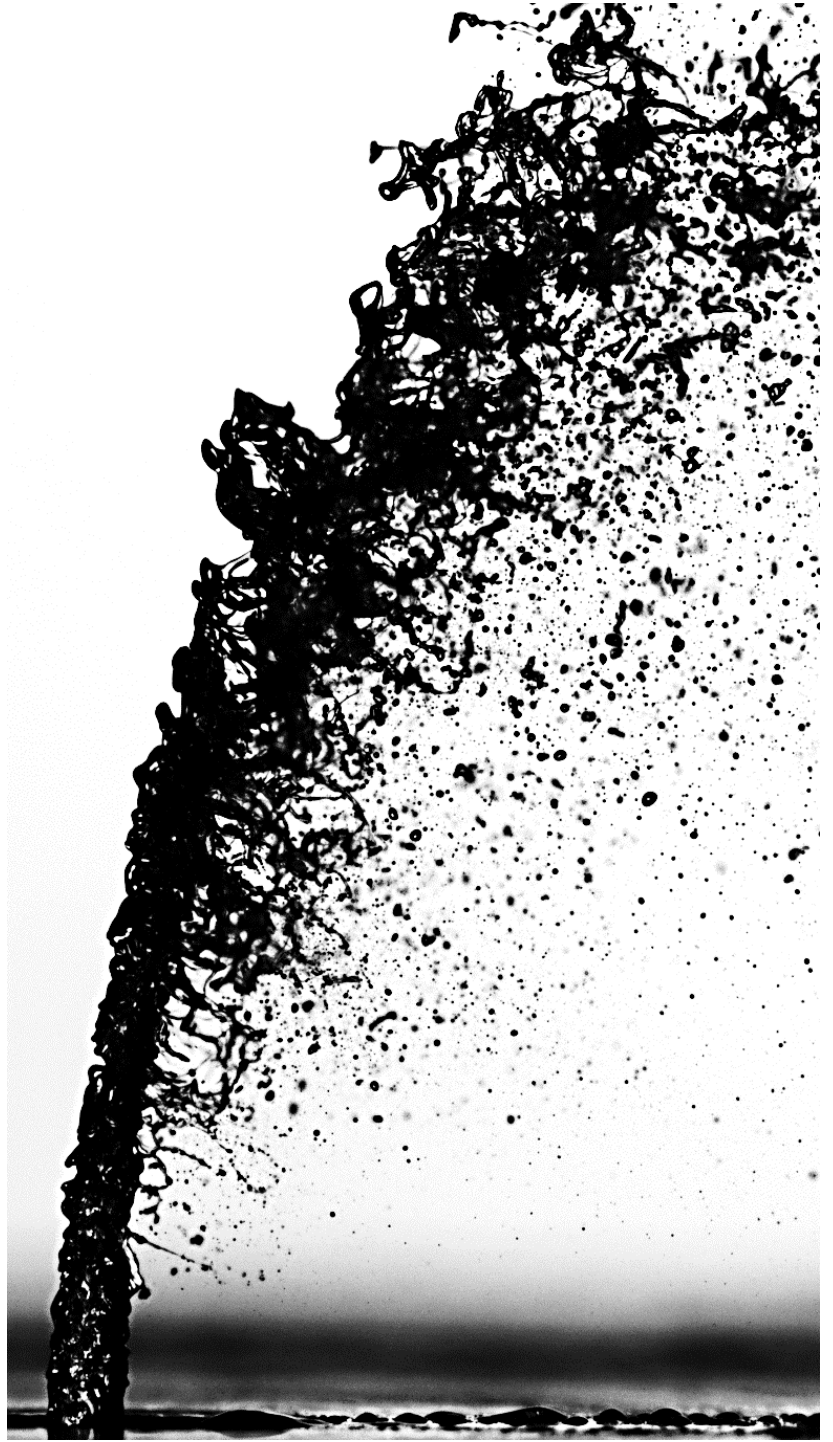

(i) J5 jet  (ii) J6 jet

**Fig. 9. Near-nozzle jet structure for J5 and J6 cases.**

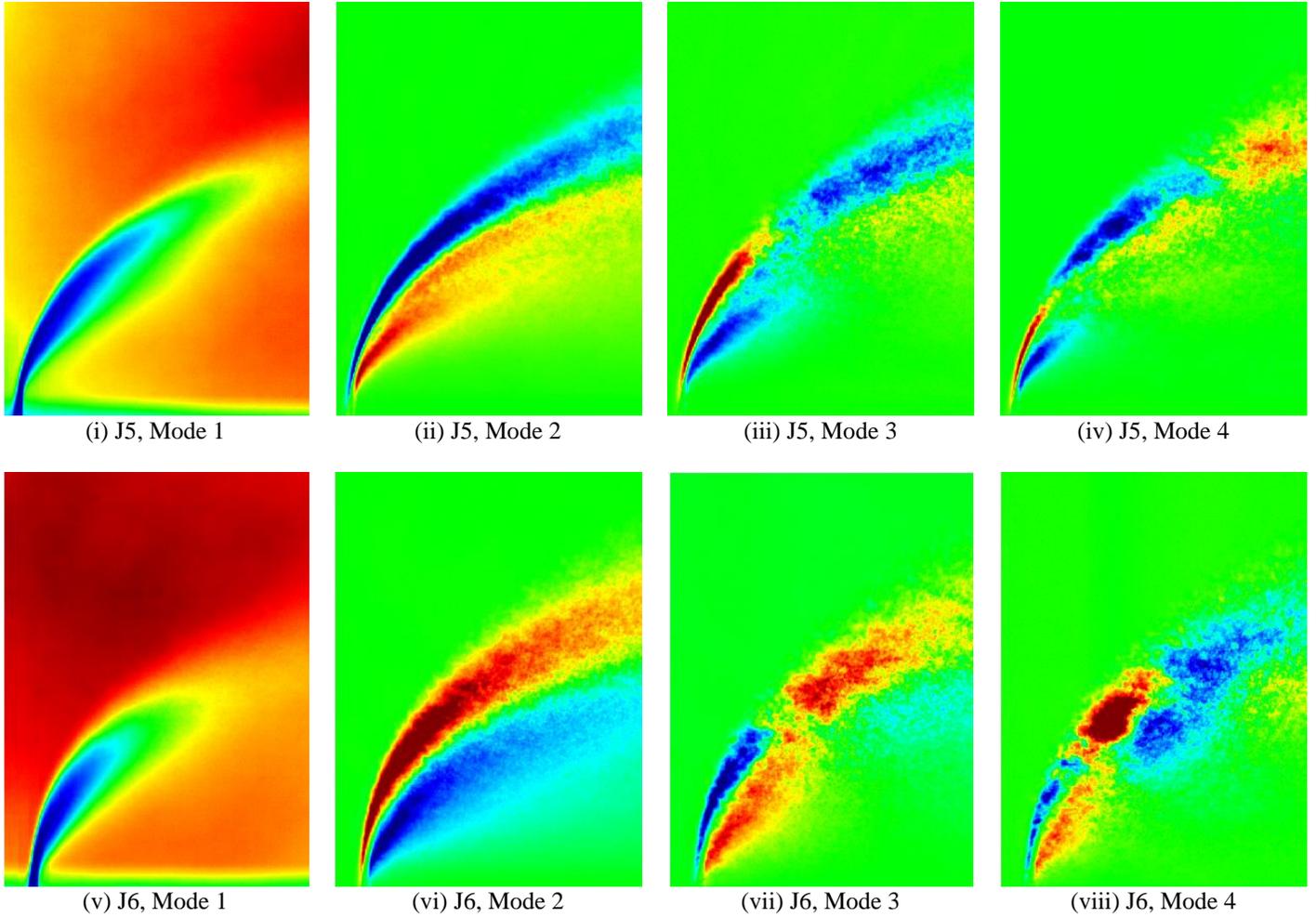

(i) J5, Mode 1  (ii) J5, Mode 2  (iii) J5, Mode 3  (iv) J5, Mode 4

(v) J6, Mode 1  (vi) J6, Mode 2  (vii) J6, Mode 3  (viii) J6, Mode 4

**Fig. 10. POD Mode shapes for J5 and J5 cases**

Figure 9 presents the near-nozzle images for case J5 and J6, and their POD and PSD modes are shown in Fig. 10 and 11 respectively. Figure 9 exhibits the stripping of droplets from the jet surface, which is expected for these conditions. However, it is evident that the stripping is more amplified for J6, and a large number of smaller droplets are observed in the lee-ward region, as compared to J5. Both jets are turbulent, but J6 appears more unstable and breaks rapidly. In Fig. 10, the mean jet trajectory appears to be shorter for J6. It is also more deflected in the crossflow direction. POD mode 2 for previous cases depicted column oscillations. However, for J5 and J6, mode 2, not only includes column fluctuations, but the stream of droplets formed by surface stripping is also included. This explains why this mode shows structures which are much wider than the column fluctuations. Also, a closer observation of PSD modes in Fig. 11 reveal that the mode 2 has a dominant peak near the zero frequency, which signifies contribution from mean flow field. This mode shape could also be interpreted to be composed of two separate streams. The higher stream consists of jet column and large ligaments following the jet trajectory. The lower stream is made up of small droplets stripped off the jet surface penetrating the cross flow. This explains why this structure in mode 2 of J6 is so wide as compared to J5, since J6 has more unstable and has higher droplet stripping, as observed in Fig. 9. This aspect becomes clearer by running the videos of high-speed images. This aspect getting captured in POD modes demonstrates the capability of POD to identify hidden features in a flow field. Mode 3 and 4 denote the jet flapping behavior in the near nozzle region, and ligament transport in the far field. Mode 3 and 4 of J5 show PSD peaks at around 1000 Hz. The same frequency is reported previously for ligament transport of cases J1 and J2. However, the ligament transport and breakup process is more complex in J6, signified by the wide range of frequencies observed in mode 4 of J6. It is interesting to note that the PSD spectrums have a more clearly defined peaks for J1 and J2, while the frequency spectrum gets wide for higher modes of J3 and J4, and wider for J5 and J6. It can be inferred that the breakup process becomes more complex with increase of *We* and $Re_l$.

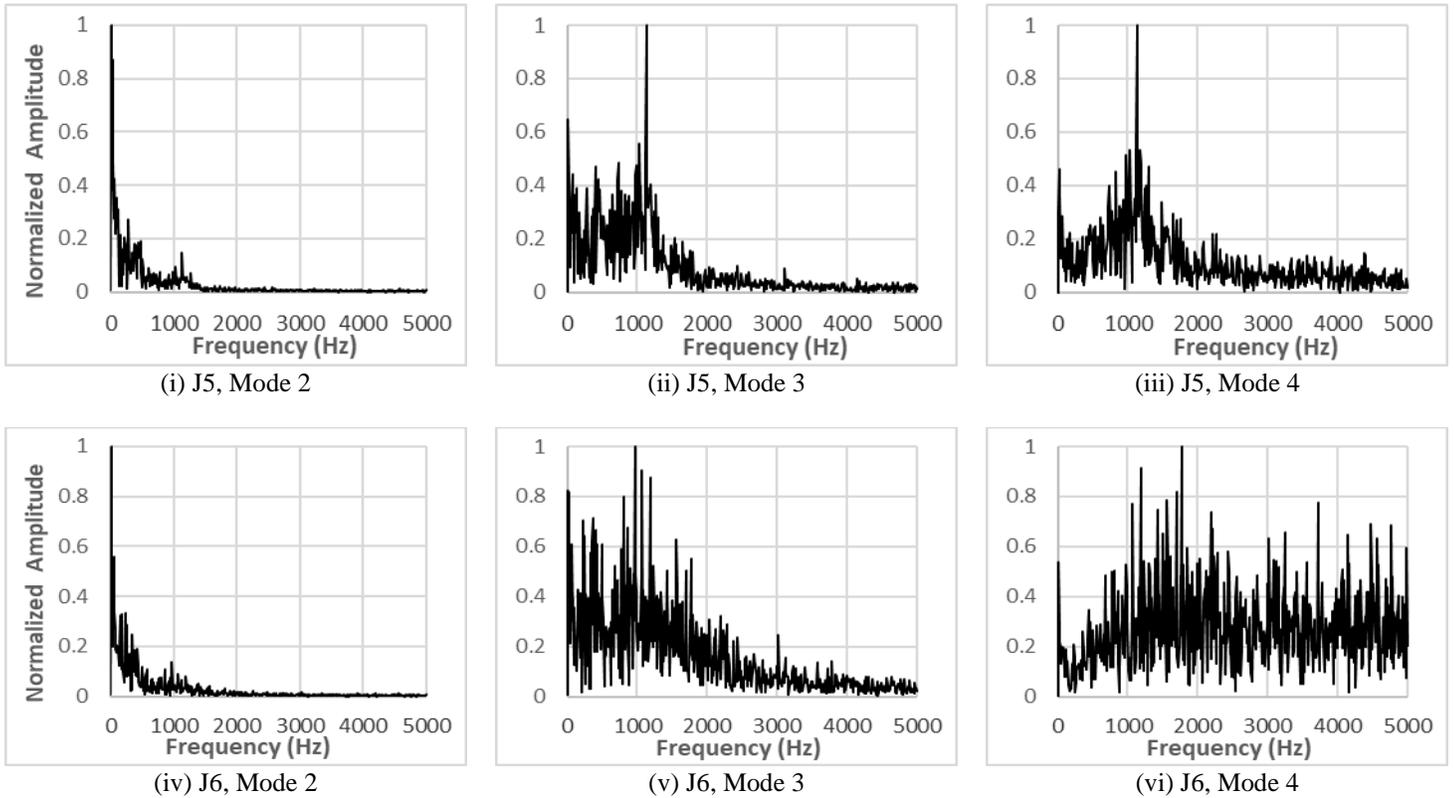

**Fig. 11 PSD spectrum for J5 and J6 cases**

## IV. Conclusion

The present study investigates liquid jet breakup in presence of crossflow of air. Liquid Reynolds number is varied from 2545 to 10180, to capture both laminar as well as turbulent jet behavior. Weber number is varied from 16 to 79, to observe different breakup modes. Injector L/D is varied from 10 to 100 to study its effect on jet breakup. High resolution, pulsed images are used to investigate the near-nozzle jet behavior. Time-resolved jet trajectory images are used to study the temporal evolution of breakup dynamics. Proper Orthogonal Decomposition of time-resolved images is used to further probe jet stability. The POD mode shapes, and PSD spectra are compared with near-nozzle images to correlate observations. It is observed that the injector ($L/D$) ratio affects the jet stability, and its eventual breakup in presence of crossflow. General observation is that jet form a higher ($L/D$) injector tends to be more unstable, and disintegrates more rapidly, its trajectory is also more deflected in crossflow direction. The POD modes and PSD spectra are also able to capture this trend. POD mode shapes show column oscillation and jet flapping occurring at higher amplitude for jet with higher ($L/D$). this observation is also verified by time-resolved videos. Sheet stripping is also amplified for higher ($L/D$). PSD spectra show increase in high frequency peaks indicating the breakup to become more complex; and POD modes to be superposition of several processes happening at different time scales with increase in *We* and *Re$_l$*. The ($L/D$) ratio is believed to alter the velocity profile of the jet which determines its stability. Future studies are planned to further probe the velocity profile and how it governs jet stability. Other geometric parameters like type of entrance, contraction ratio, etc. need to be investigated in detail to improve understanding of the breakup phenomenon.

## Acknowledgments

The author gratefully acknowledges support provided by R. V. Ravikrishna (IISc Bangalore) and the Aeronautics Research and Development Board, Government of India, for the experimental part of this work.